\documentclass[floatfix,prb,twocolumn,showpacs]{revtex4}
\usepackage{dcolumn}
\usepackage{graphicx}
\usepackage{amsmath,epsf}
\begin{document}
\author{S. Elgazzar,$^{1,2}$ I. Opahle,$^1$ R. Hayn,$^{1,3}$ and P. M. Oppeneer$^{1,4}$}
\affiliation{
$^1$Leibniz-Institute of Solid State and Materials Research, P.O. Box 270016,~D-01171 
Dresden, Germany\\
$^2$Department of Physics, Faculty of Science, Menoufia University,
Shebin El-kom, Egypt\\
$^3$Laboratoire Mat\'eriaux et Micro\'electronique de Provence, Case 142,
Universit\'e d'Aix-Marseille III, Facult\'e Saint-Jer\^ome, 13397 Marseille
Cedex 20, France \\
$^4$Department of Physics, Uppsala University, Box 530, S-751 21 Uppsala, Sweden}
\title{Calculated de Haas-van Alphen quantities of Ce$M$In$_5$ ($M=$Co, Rh, and Ir)
 compounds}
\date{\today}
\begin{abstract}
We report a critical analysis of the electronic structures and 
de Haas-van Alphen (dHvA) quantities of the heavy-fermion superconductors
CeCoIn$_5$, CeRhIn$_5$, and CeIrIn$_5$. The electronic structures are 
investigated {\it ab initio} on the basis of full-potential band-structure
calculations, adopting both the scalar- and fully relativistic
formulations within the framework of the local spin-density approximation
(LSDA). In contrast to another recent study, in which a pronounced 
change of the Fermi surface due to relativistic effects and therefore the
importance of relativistic interactions for the superconductivity was claimed, 
we find only minor relativistic modifications of the band structure
in our calculations.
The {\it ab initio} calculated dHvA quantities are in good agreement with
experimental data for CeCoIn$_5$ and CeIrIn$_5$, when we adopt the 
delocalized LSDA description for the Ce $4f$ states. For CeRhIn$_5$,
however, a better agreement with experiment is obtained when the Ce
$4f$ electron is treated as a localized core electron. The implications for 
an emerging picture of the localization behavior of the $4f$ electron in these
materials 
are discussed. We furthermore
compare our calculated dHvA quantities with other recent relativistic 
calculations and discuss the differences between them.
\end{abstract}

\maketitle

\section{Introduction}

A new group of fascinating heavy-fermion
superconductors that crystallize in the HoCoGa$_5$ structure
was discovered a few years ago.\cite{hegger00}
To the superconductors of this group that have received wide attention
belong CeCoIn$_5$, CeRhIn$_5$, and CeIrIn$_5$
(Refs. \onlinecite{hegger00,petrovic01a,petrovic01b}).
CeCoIn$_5$ and CeIrIn$_5$ are superconductors at ambient pressure,
\cite{petrovic01a,petrovic01b}
with $T_c= 0.4$ and 2.3~K, respectively, whereas CeRhIn$_5$
becomes a superconductor (with $T_c =2.1$~K) under pressure.
\cite{hegger00}
At ambient pressure CeRhIn$_5$ orders antiferromagnetically
with an incommensurate spin spiral below the N\'eel temperature $T_N = 3.9$ K
(Ref. \onlinecite{bao00}).
The entanglement of heavy-fermion behavior, antiferromagnetism, and
superconductivity seems to point to an unconventional pairing
mechanism.\cite{petrovic01a,petrovic01b} Unambiguous evidence for
unconventional spin-singlet $d$-wave superconductivity in CeRhIn$_5$ and CeCoIn$_5$
was recently provided by a number of experiments.
\cite{zheng01,movshovich01,izawa01,ormeno02}
The nature of the pairing mechanism, however, could not yet be established.

An important issue which has become intensively debated is the role of the Ce $4f$
electrons and their degree of localization. The high specific heat coefficients
\cite{petrovic01a,petrovic01b}
of about 300 mJ/molK$^2$ to 750 mJ/molK$^2$ for CeCoIn$_5$ and CeIrIn$_5$,
respectively, indicate at least some $4f$ delocalization at ambient pressure.
The $4f$ localization has been studied in some detail by de Haas-van Alphen (dHvA)
experiments,
\cite{cornelius00,hall01,alver01,hall01b,settai01,haga01,shishido02,shishido02b,shishido03}
in which especially CeRhIn$_5$ received attention.
\cite{cornelius00,hall01,alver01,hall01b}
Although the dHvA data of pure
CeRhIn$_5$ were initially interpreted using an itinerant band-structure
calculation with delocalized $4f$ electrons,\cite{hall01} subsequent
measurements of the dHvA quantities of
Ce$_x$La$_{1-x}$RhIn$_5$ indicated \cite{alver01} that the Ce $4f$ electrons
remain localized for all $x$. The rather low specific heat coefficient
$\gamma \approx 50$ mJ/molK$^2$ at ambient pressure would corroborate localized
$4f$ behavior in CeRhIn$_5$. Even in the localized picture, 
the $4f$ electrons are responsible for the magnetism which could, for
example, through spin degrees of freedom, give rise to an unconventional
pairing interaction.

De Haas-van Alphen measurements in connection with band-structure calculations
are a very useful tool to analyze a multiband situation with a complicated
Fermi surface. Such studies have been performed for the three Ce-115 compounds in
question by several groups with contradictory results, however. In the
case of CeCoIn$_5$, the authors of Ref.\ \onlinecite{settai01} found a
reasonable agreement of the calculated dHvA data with the measured values,
\cite{hall01b,settai01} having in their band structure three bands 
crossing the Fermi surface (see also
Ref.\ \onlinecite{costa03}). A similar agreement was found for the 
superconducting compound CeIrIn$_5$ (see Ref.\ \onlinecite{haga01}).
Conversely, another recent study \cite{maehira03} claimed an important influence
of relativistic effects on the electronic structure of these two compounds, 
leading to a fourth band crossing the Fermi surface which would give rise to
additional dHvA frequencies. If relativistic effects are important
then this would obviously affect the microscopic understanding of the 
superconductivity (see, e.g., Ref.\ \onlinecite{hotta03}) and possibly
of the pairing mechanism, which remains an unresolved issue.
Another controversy was already noted and concerns CeRhIn$_5$ 
which is antiferromagnetic at ambient pressure. The authors of Ref.\
\onlinecite{hall01} interpreted its dHvA data by a nonmagnetic itinerant band
structure calculation treating the $4f$ electrons as valence states and stating
a reasonable agreement. On the contrary, the localized nature of the $4f$
electrons in CeRhIn$_5$ was proposed in Ref.\ \onlinecite{shishido02} 
on the basis of a comparison to the dHvA data of LaRhIn$_5$. 
Localization of the $4f$ electron was also revealed \cite{alver01}
by dHvA investigations of Ce$_x$La$_{1-x}$RhIn$_5$,
but no corresponding band-structure calculation was performed up to now. 

Taking into account the apparent similarity within the electronic structure of
all three compounds, a comparative and critical analysis of the electronic
structures and de Haas-van Alphen quantities is highly desirable. 
Also, performing a complete theoretical analysis of the dHvA data for 
$H\parallel c$ of the three Ce-115 compounds using the same
numerical method (as will be presented here) allows one to critically
assess the reliability of former calculations. 
Comparing scalar-relativistic and fully relativistic
calculations will answer the discussion concerning the importance of
relativistic contributions, started in Ref.\
\onlinecite{maehira03}. In addition,
comparing the Co-, Ir-compounds on one side with
the Rh-compound on the other side (treating the $4f$ electron as core 
electron or not) may possibly prove whether the $4f$ electrons are localized 
or not.

For our study we use the full potential local orbital (FPLO) method 
\cite{koepernik99,opahle01} for all three
compounds CeCoIn$_5$, CeRhIn$_5$, and CeIrIn$_5$. We calculate the extremal
Fermi surface cross sections and effective masses and restrict ourselves to
$H \parallel c$ for which most experimental data exist. We compare 
scalar-relativistic and fully relativistic calculations, and, in the case of
CeRhIn$_5$, the localized and the delocalized descriptions of Ce $4f$
electrons. After presenting our method (Sec.\ II) and a short discussion of
the resulting band structures (Sec.\ III) we describe the extremal orbits in
Sec.\ IV. The critical comparison with former calculations and experimental
results (Sec.\ V) allows to draw our conclusions in Sec.\ VI.

\section{Computational Method}

We performed band-structure calculations using
both the scalar-relativistic and the fully relativistic versions of the
full potential local orbital minimum-basis band-structure method.
\cite{koepernik99,opahle01} In these calculations, the following basis sets
were adopted for the valence states: 
the $ 4f;5s5p5d;6s6p$ states for Ce when the $4f$ are treated as
valence states, while for Co, Rh, Ir and In were used  $3s3p3d;4s4p$, 
$4s4p4d;5s5p$,  $5s5p5d;6s6p$ and $4d;5s5p$ respectively. The high lying $5s$
and $5p$ semicore states of Ce, which might hybridize with the $5d$ 
valence states, are thus included in the basis. The compression parameters
$x_0$ were optimized for each basis orbital separately by minimizing the total
energy. 
For the site-centered
potentials and densities we used expansions in spherical harmonics up
to $\it {l_{max}}$=12. The number of $k$-points in the irreducible part of
%pmo
Brillouin zone was 726, but some pivotal calculations were made also with 
higher numbers of $k$-points.
%pmo
The Perdew-Wang \cite{perdew92} parameterization of the
exchange-correlation potential in the local spin-density
approximation (LSDA) was used. 

The dHvA cyclotron frequency $F$---which is proportional to the Fermi surface
cross section---and the cyclotron mass $m$ of the extremal
orbits are calculated numerically by discretizing the Fermi velocities
on $k$-points along the
orbit and by a subsequent Romberg integration.\cite{oppeneer87}

\section{Band structure}

The compounds Ce$M$In$_5$ ($M$=Co, Rh, and Ir) crystallize in the tetragonal
structure, space group P4/mmm (space group number 123) and are built of
alternating stacks of CeIn$_3$ and $M$In$_2$. We performed nonmagnetic band
structure 
calculations for the experimental lattice parameters, which are
(in atomic units):
$a=8.714$ $a_0$ and $c=14.264$ $a_0$ (CeCoIn$_5$), $a=8.791$ $a_0$ and $c=14.252$
$a_0$ (CeRhIn$_5$), and $a=8.818$ $a_0$ and $c=14.205$ $a_0$ (CeIrIn$_5$). 
For the one special In-position we also adopted the experimental values.

\begin{figure}%[fig1.epsi]
%p \includegraphics[angle=-90,width=8cm]{Fig1-CeCoIn5-4fv.epsi}
\includegraphics[angle=-90,width=8cm]{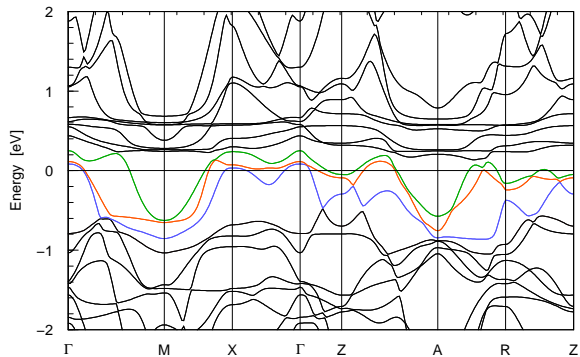}
\caption{(Color online) 
The energy bands of nonmagnetic CeCoIn$_5$ calculated using the fully
relativistic FPLO method. The bands 131, 133, and 135 that cut the Fermi
energy are highlighted by the colors.}
\label{fig1}
\end{figure}

The band structure of CeCoIn$_5$ computed with the fully relativistic scheme 
is presented in Fig.\ \ref{fig1}. The band structure of CeCoIn$_5$ obtained
with the scalar-relativistic scheme, as well as 
the band structures of nonmagnetic CeRhIn$_5$ and CeIrIn$_5$ are only slightly 
different and therefore not presented here. We find for each of the Ce-115 systems   
using either the scalar-relativistic or fully relativistic scheme
three bands which 
cross the Fermi surface. These bands are denoted as Band 131, Band 133 and Band 135, 
according to their number in the valence band complex of the fully relativistic
calculation counted from below. Following the line $\Gamma-M$ in Fig.\ \ref{fig1} the
crossing points of Bands 131, 133 and 135 (in that order) with the Fermi level
are clearly visible. The influence of relativistic effects is most pronounced 
along the $\Gamma-Z$ line, where the upper two bands 133 and 135 remain 
degenerate in the scalar-relativistic 
calculation. However, the fully relativistic effect along $\Gamma - Z$ and at
other places in the Brillouin zone (BZ) have only minor influence on the
Fermi surface. We find only small differences in the extremal Fermi surface
cross sections which will be discussed in the next Section.

\begin{figure}%[fig2.epsi]
%p \includegraphics[angle=-90,width=8cm]{Fig2-CeRhIn5-4fc.epsi}
\includegraphics[angle=-90,width=8cm]{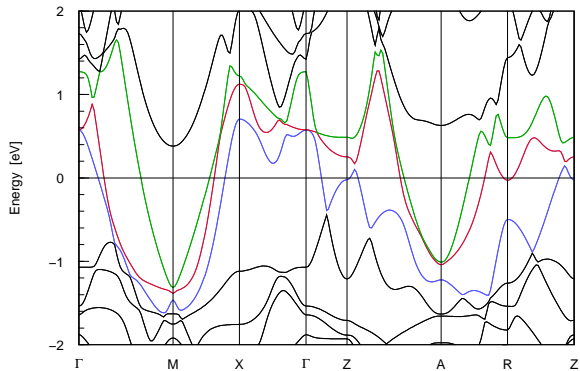}
\caption{(Color online)
The energy bands of CeRhIn$_5$ calculated with the Ce $4f$ electron 
treated as a core electron. The important bands $131-135$ are colored
as in Fig.\ \protect\ref{fig1}.
}
\label{fig2}
\end{figure}

Also, the differences in the band structures calculated for the three compounds 
CeCoIn$_5$, CeRhIn$_5$, and CeIrIn$_5$ are small. For example, 
in the fully relativistic calculation of
CeRhIn$_5$, band number 133 does not cross the Fermi level between $A$ and $R$
in variance to Fig.\ 1, where it just crosses the Fermi level.
As a result, the two extremal orbits (for $H
\parallel c$) of band 133 which are centered around the $A$ point for
CeCoIn$_5$ are no longer closed orbits for CeRhIn$_5$. Instead, one finds for
CeRhIn$_5$ corresponding orbits which are closed around other points in
$k$-space, however.

CeRhIn$_5$ differs in its physical properties from the other two
compounds and a debate concerning the localization of the Ce 4$f$ electron 
has emerged.\cite{hall01,shishido02,alver01} Below we will see that
the agreement between calculated and measured dHvA frequencies is indeed
significantly worse for CeRhIn$_5$ (in comparison to the Co- and Ir-compounds)
if we treat the Ce $4f$ electrons itinerantly. Therefore, we present in Fig.\
\ref{fig2} the result of a fully relativistic band-structure calculation for
CeRhIn$_5$, where its Ce $4f$ electron was treated as a core electron
($4f$-c scheme). By 
comparison with Fig.\ 1 one can easily conclude that the seven, 
nearly  dispersionless
bands in Fig.\ 1, situated at about 0.5 eV are of Ce $4f$ character. These
bands are especially clearly visible above the Fermi level at $A$ and $M$,
but they are mixed with other bands elsewhere and the lowest one drops below
the Fermi level between $\Gamma$ and $X$. 
%R
By calculating additionally the atomic character within each band (not shown)
we have found that the bands which
cross the Fermi level in Fig.\ 1 have a considerable $f$ character
as well. Especially
pronounced is the $f$ character at the Fermi energy between $X$ and $\Gamma$,
as well as between $R$ and $Z$. Considerable $f$ character below the Fermi
energy exists only at the high-symmetry points $Z$ and $R$. Other states
which contribute significantly at the Fermi energy are the In $5p$ orbitals.
This is especially true for the bands centered around the $A$ and $M$ points. 
If we compare the energy bands in Figs.\ 1 and 2 between $M$ and $\Gamma$, 
or between $M$ and $X$, 
we find in both cases three bands crossing the Fermi energy. The two upper
bands (bands 133 and 135 in the fully relativistic case) give rise to closed
orbits centered at $M$ which are of predominantly In $5p$ character and
%pmo present irrespective of the treatment of the Ce $4f$ electrons.
present at $E_F$ irrespective of the treatment of the Ce $4f$ electrons.
%pmo
However, the areas are different. The same is true
for the two, electron-like bands, centered around the $A$ point. 
In the present study, we did not try to interpret the dHvA data of CeRhIn$_5$
with a band-structure calculation taking into account the correct
low-temperature antiferromagnetic order of that compound.

Let us shortly compare our band-structure results with those published already
in the literature. We find that 
our FPLO result agrees best with the FLAPW band structure
\cite{haga01} published for CeIrIn$_5$. This agreement concerns also the corresponding
Fermi surfaces and, for CeCoIn$_5$ it holds as well (see Refs.\
\onlinecite{settai01} and \onlinecite{shishido02}). The relativistic linear
augmented-plane-wave (RLAPW) calculations published in Ref.\
\onlinecite{maehira03} show in contrast to our results a fourth band
crossing the Fermi surface. The corresponding band in our calculation
can be seen in Fig.\ 2 at about
0.4 eV {\it above} $E_F$ at the $M$ point and it consists mainly of In $5p$ character. 
It could drop below the Fermi level due 
to the use of a non-optimal basis set. The
nonrelativistic band structure of CeCoIn$_5$ published in Ref.\
\onlinecite{costa03} shows a rough agreement with our results with respect to
the band characters, but considerable differences due to relativistic effects
exist near $E_F$. For instance, due to the lack of spin-orbit splitting, the $4f$
band complex is much narrower there.\cite{costa03}
Finally, band-structure results and the corresponding Fermi surface of CeRhIn$_5$ 
were reported in Ref.\
\onlinecite{hall01}. Whereas the band structure is quite similar to
our results, there exist some differences in the calculated dHvA data as well as
in the assignment of the measured dHvA frequencies. 

\section{Extremal orbits}

Based on the band-structure results, we calculated the dHvA frequencies and
effective masses for all extremal orbits with $H\parallel c$. The most relevant
calculations were performed in the fully relativistic scheme for which a
complete analysis has been performed and which are compared with the 
scalar-relativistic calculations. 
The individual extremal orbits were already shown in  Fig.\ 5 of 
Ref.\ \onlinecite{haga01} (see also Refs.\ \onlinecite{settai01} and \onlinecite{shishido02})
which we therefore 
%R do 
not repeated here. 
For clarity sake we also use the same notation\cite{haga01} except for
one minor  modification: in some cases the orbit $a$ does not exist and is
broken into two other orbits which we then denote as $a_1$ and $a_2$. 
The band numbers in Tables I--III refer to the valence bands in the fully 
relativistic scheme counted from below as introduced in the previous Section. 
In the case of CeRhIn$_5$ we present also some dHvA data for the most prominent 
Fermi surface sheets which we calculated
using the $4f$-c scheme. We did, however, not attempt to identify all the 
experimentally observed orbits.
Lastly, we note that some orbits in either the scalar- or fully relativistic
scheme are not presented because they do not exist as a closed, extremal
orbit.
 
\begin{table}[ht]
\caption{Calculated dHvA frequencies $F$ (in kT) and effective masses 
$m$ (in $m_0$)
for CeCoIn$_5$ with $H\parallel c$. The calculations were performed treating
the Ce $4f$ electrons as delocalized and adopting either
the scalar-relativistic or the fully relativistic approach. 
The notation
of the extremal orbits can be found in Ref.\ \onlinecite{haga01}.} 
\begin{ruledtabular}
\begin{tabular}{cccrrrr} 
%p \hline \hline
      &central &  band          &\multicolumn{2}{c}{$F$ (kT)}
&\multicolumn{2}{c}{$m~(m_0)$}\\
\cline{4-5}\cline{6-7}
symbol	& point & no.&sc-rel.        &rel.   &sc-rel.        &rel.\\
\hline 
$g$  &$\Gamma$		&131	&0.809	&0.761  	&-0.661	& -0.814\\ 
$h$  &$X$	                &131	&0.460	&0.438  	&-0.669
&-0.974 \\ 
$\beta{_1}$     	&$M$	&133	&12.910 	&12.680 	&
2.279 	& 2.314   \\ 
$\beta{_2}$     	&$A$	&133	&6.312  	&6.295  	&1.232
& 1.474    \\ 
$c$               	&$A$	&133	&13.496	       &13.113 	&-3.178
&-3.645 \\ 
$\alpha_1$     &$A'_{k_z=0.18}$ & 135	&5.399       	&5.352  	& 1.543
&1.621\\
$\alpha_2$   	&$M$	&135	&4.599   	& 4.475 	&0.977
&0.996    \\ 
$\alpha_3$   	&$A$	&135	&4.069   	& 4.060 	&1.155
&1.317    \\ 
$a_1$                	&$Z$	&135	&1.280  	&1.264
&1.445  	&1.536    \\ 
$a_2$                	&$R$	&135	& 1.186 	&1.136
&0.909  	& 0.946    
%p \hline  \hline
\end{tabular}
\end{ruledtabular}
\label{table1}
\end{table}
\begin{table}[ht]
\caption{Calculated dHvA frequencies $F$ and effective masses $m$ for $H\parallel c$
in CeIrIn$_5$.} 
\begin{ruledtabular}
\begin{tabular}{cccrrrr} 
%p \hline \hline
       &central &band
&\multicolumn{2}{c}{$F$ (kT)}&\multicolumn{2}{c}{$m~(m_0)$}\\ 
\cline{4-5}\cline{6-7}
symbol &point&no.&sc-rel.        &rel.   &sc-rel.        &rel.\\ \hline
$g$       	&$\Gamma$&131	&0.701    	&0.678  	&-0.566  &
-0.632	\\ 
$h$             &$X$&131	&0.137  	&0.224  	& -0.776
&-0.693	\\ 
$\beta_1$    	&$M$&133	&12.086 	&12.106 	&1.957  	&
2.226	\\ 
$\beta_2$    	&$A$&133	&6.083   	&5.963  	&1.062  	&
1.156  	\\ 
$c$                     	&A&133	&13.175  	&12.772 	&-2.860
&-3.386 	\\ 
$\alpha_1$&$A'_{k_z=0.18}$	&135	&4.994 	&4.984  	&1.387	&1.694
\\
$\alpha_2$  	&$M$&135	&4.148   	& 4.017 	&0.929   	&1.026
\\ 
$\alpha_3$  	&$A$&135	&3.877  	&3.929  	&0.998  	&1.324
\\
$a_1$ 	&$Z$&135	&1.418    	&open   	&1.903  	&open
\\ 
$a_2$ 	&$R$&135	&1.262 	&open   	&1.150  	&open 	\\ 
$a$                     	&$A$&135	&open   	&15.011 	&open
&-4.527   
%p \hline \hline
\end{tabular}
\end{ruledtabular}
\label{table2}
\end{table}

Let us now characterize the various extremal orbits in more detail. 
{\bf Band 131} gives rise to two small orbits around $\Gamma$ (orbit $g$) and 
around
$X$ (orbit $h$). These two orbits are hole-like and they are present in all
three compounds. {\bf Band 133} creates the electron-like orbit $\beta_1$ 
around $M$, and also 
two orbits, $\beta_2$ and $c$, around the $A$ point which are electron- and
hole-like, respectively. The two distinct kinds of orbits around $A$ may be understood
from the band structure (Fig.\ 1), since Band 133 crosses the Fermi level
twice along each of the high-symmetry directions $A - R$ and $ A -Z$, respectively. 

The most important orbits are $\alpha_1$, $\alpha_2$ and $\alpha_3$, which are
connected with {\bf Band 135}. These orbits are the ones that are most clearly
visible in experiment\cite{shishido02,shishido02b,shishido03}
(together with $\beta_2$) and, in addition, 
they were observed in all three compounds. They
correspond to a cylindrical Fermi surface sheet along the $ M-A$ axis in the BZ. 
Orbit $\alpha_3$ is centered around the $A$ point, orbit $\alpha_2$ around $M$
point, and orbit $\alpha_1$ is centered
around a point in between $A$ and $M$. These three orbits are all electron-like.
The hole-like orbit $a$ around point $A$ in $k$-space exists only in the
fully relativistic calculation of CeIrIn$_5$. In all other cases it breaks up 
into two orbits (for which we introduced the notations $a_1$ and $a_2$)
which are centered around $ Z$ and $R$, respectively, and which are electron-like.

To explain the peculiarities of the three compounds summarized in Tables I--III
some comments are given in the following:
\begin{itemize}
%pmo \item{Orbit $\alpha_1$ exists also in the scalar-relativistic calculations of
%pmo all three compounds, but was not calculated. [{\bf Roland: it would be good if
%pmo this was just done, I think}]}
\item{In the case of CeRhIn$_5$ (Table III), the orbit $\beta_2$ is an extremal
orbit around the $A$ point in the scalar-relativistic scheme as well as
in the $4f$-c scheme. In the
fully relativistic case, the corresponding orbit $\beta_2$ has an extremal
cross section at $k_z=0.25$ (denoted by $A'$) and it disappears for larger
values of $k_z$ including the $A$ point.}  
\item{For CeRhIn$_5$ (Table III), the extremal orbit $c$ is centered around $A$
in the scalar-relativistic scheme, but has an extremal cross section around $A'$
with $k_z=0.27$ in the fully relativistic case.}
\item{In the $4f$-c scheme (only relevant for CeRhIn$_5$, Table III) one can find
more extremal orbits than given in the Table. Only those extremal orbits are
presented which have a clear correspondence in the fully relativistic
scheme. Also, there are extremal orbits in the fully relativistic
calculation which have no correspondence in the $4f$-c scheme.}
\item{As was already noted, the orbit $a$ exists only in the fully relativistic
calculation for CeIrIn$_5$ (Table II) but breaks up into two orbits $a_1$
and $a_2$ in the scalar-relativistic calculation for that compound as well
as for the other Ce-115 compounds.} 
\end{itemize}

\begin{table*}[ht]
\caption{De Haas-van Alphen frequencies $F$ and effective masses $m$ calculated for 
CeRhIn$_5$ with $H\parallel c$. The presented dHvA quantities were calculated
treating the Ce $4f$ electrons as delocalized, adopting either the
scalar-relativistic or the fully relativistic scheme, as well as with
the $4f$-c scheme, in which the Ce $4f$ electron is treated as localized.
}
\begin{ruledtabular}
\begin{tabular}{cccrrcrrc} 
         &central 	&band &\multicolumn{3}{c}{$F$ (kT)}    &\multicolumn{3}{c}{$m~(m_0)$}\\
	 \cline{4-6}\cline{7-9}
symbol   &point	        &no.&sc-rel.        &rel.       &$4f$-c	&sc-rel.&rel.&$4f$-c \\ 
\hline
$g$   	&$\Gamma$	&131	&0.848	&0.821    	&	& -0.645	&-0.807&	\\
$h$            	&$X$	&131	&0.544 	&0.537  	&	&-0.666 	&-0.899&	\\
$\beta_1$      	&$M$	&133	&12.759 	&12.201	&10.250	&2.007  	&1.994&0.629	\\
$\beta_2$       &$A$	&133	&5.808 	& open   	&6.073	&1.080 	&open&0.551	\\
$\beta{_2}$ &$A'_{k_z=0.25}$	&133	&         	&6.534   	&	&	&1.484&		\\
$c$            	&$A$	&133	&12.976 	& open   	&	&-2.707  &open&		\\
$c$         &$A'_{k_z=0.27}$	&133	&	&15.684	&       	& &-5.970	&	\\
$\alpha{_1}$&$A'_{k_z=0.19}$	&135	&	&5.383 	&4.567   	&		&1.535&0.549	\\
$\alpha{_2}$    	&$M$	&135	&4.371 	& 4.213	&3.928	&0.894   	&0.952&0.467	\\
$\alpha{_3}$    	&$A$	&135	&4.052  	&3.899  	&3.483	&0.997 	&1.203&0.472	\\
$a_1$                	&$Z$	&135	&1.414 	&1.312   	&	&1.310	&1.524&	\\     
$a_2$                	&$R$	&135	&1.140  	&1.051  	&	&0.836	&0.955&	
\end{tabular}
\end{ruledtabular}
\label{table3}
\end{table*}

\section{Discussion and comparison}

In Table \ref{table4}  we compare our calculated dHvA frequencies for {\bf
CeCoIn$_5$} to the available experimental data.
Our values---calculated assuming delocalized $4f$'s---agree very well with 
the theoretical results reported  by
Settai {\it et al.} \cite{settai01} (who give, however, not all frequencies that
exist). The two published experimental data sets
\cite{settai01,hall01b} agree reasonably well with one another, too, and with the
theoretical values. The experimental data of both groups\cite{settai01,hall01b} complete each
other to some extent. Taking all 
results together, only orbits $c$, $a_1$, and $a_2$ were never observed. It
might be that orbit $a$ is not split in reality and, consequently,
it should be expected at a high frequency. In general it is increasingly
difficult to detect high frequency
orbits with good precision. This difficulty of the dHvA technique could also be the reason
why orbit $c$ has not been detected.

\begin{table}[ht]
\caption{Comparison of several experimental 
%p (Refs. \onlinecite{settai01} and \onlinecite{hall01b})
and theoretical dHvA frequencies $F$ (in kT) for CeCoIn$_5$.} 
\begin{ruledtabular}
\begin{tabular}{cccccc}
symbol&symbol&\multicolumn{2}{c}{theory $F$}&\multicolumn{2}{c}{experiment 
$F$} 
 \\   
 \cline{3-4}\cline{5-6}
Refs.\ \onlinecite{settai01,haga01} &Ref.\ \onlinecite{hall01b} &present &
Ref.\ \onlinecite{settai01} &Ref.\ \onlinecite{hall01b} &Ref.\
\onlinecite{settai01} \\ \hline 
$c$                     	&		&13.11  	& 13.30  	&
 &\\  
$\beta{_1}  $  	&        		&12.68  	& 13.00 	&
 &12.00 \\ 
$\beta{_2}  $  	& F$_6$  	&  6.30  	&6.45 	&7.535 	& 7.50\\  
$\alpha{_1}$  	& F$_5$  	& 5.35   	& 5.43 	& 5.401 	& 5.56
 \\ 
$\alpha{_2}$  	& F$_4$  	& 4.48  	& 4.53 	& 5.161 	&
 4.53\\  
$\alpha{_3}$  	& F$_3$  	&4.06    	&  3.90 	& 4.566
 &4.24\\  
$a{_1}$        	&       		&1.26     	&
 &	&\\ 
$a{_2}$              	&       		&1.14     	&
 &	&\\ 
$g$                     	&F$_2$   	&0.76    	&         	&
 0.411	&\\ 
$h$                     	&F$_1$   	& 0.44    	&        	&0.267
 &
\end{tabular}
\end{ruledtabular}
\label{table4}
\end{table}

The situation is similar for {\bf CeIrIn$_5$} (see Table \ref{table6}). The
agreement of our data with the FLAPW results \cite{haga01} is excellent. 
The correspondence with the experimental data is also gratifying.
Nearly all the predicted orbits were experimentally observed. Only the 
orbits $g$, $a$, and $c$ are missing. The later ones belong to the more
difficult to detect high-frequency orbits, but $g$ is a low-frequency orbit.
It could be that experimentally the two orbits $g$ and $h$ could not be separated or
that the intensity of $g$ was too low.
The orbits $\beta_2$ and $\alpha_1$ displayed some splitting in the experiment.
\cite{haga01} Despite the good agreement achieved for the dHvA frequencies, the experimental
cyclotron masses are extremely enhanced compared to the calculated (unehanced)
orbital masses. This experimental observation understandably
reflects the heavy fermion behavior of CeIrIn$_5$, which has a specific heat
coefficient 
\cite{petrovic01a}
of about 750 mJ/molK$^2$.
A similar mass enhancement exists for CeCoIn$_5$ (see Ref.\
\onlinecite{settai01}).
%R
The mass enhancement is connected with the 4$f$ character of the
bands. 

\begin{table}[ht]
\caption{Comparison of experimental and theoretical dHvA frequencies and masses
of CeIrIn$_5$.}
%\begin{tabular}{crrcrrc} \hline \hline
\begin{ruledtabular}
\begin{tabular}{ccccccc} 
symbol&\multicolumn{2}{c}{theory $F$}&exp.\
$F$&\multicolumn{2}{c}{theory  
$m$}&{exp.\ $m$}\\
\cline{2-3}\cline{5-6}
Ref.\ \onlinecite{haga01}		&present	&Ref.\
\onlinecite{haga01} & Ref.\ \onlinecite{haga01} &present	&
Ref.\ \onlinecite{haga01} & Ref.\ \onlinecite{haga01} \\ \hline 
$a$	       	&15.01  	&15.20  	&                  	&4.53
&4.89	&\\  
$c$	       	&12.77  	&12.80  	&                  	&3.39
&3.95	&\\ 
$\beta{_1}$    	&12.11  	&12.40  	&12.00         	&2.23	&3.32
&32\\  
$\beta{_2}$    	&5.96    	&6.18    	&6.11-6.59  	&1.16	&1.33
&21-30\\ 
$\alpha{_1}$  	& 4.98   	&5.00    	&5.07-5.56  	&1.69	&1.89
&17-25\\  
$\alpha{_2}$  	& 4.02   	&4.03    	&4.53	  	&1.03	&1.15
&29\\ 
$\alpha{_3}$  	&3.93    	&3.65    	&4.24		&1.32	&1.37
&10\\ 
$g$  	&0.68    	&0.70    	&
&0.63	&0.64	&\\ 
$h$  	&0.22    	&0.14    	&0.27
&0.69	&1.04	&6.3
\end{tabular}
\end{ruledtabular}
\label{table6}
\end{table}

The compound {\bf CeRhIn$_5$} exhibits a remarkably different behavior. The
dHvA experiments performed in the antiferromagnetic state
\cite{hall01,shishido02} detected a multitude of low-frequency branches
besides the main branches $\beta_2$, $\alpha_1$ and $\alpha_{2,3}$. Here, our 
focus is not to describe this multitude of low-frequency orbits, for which
most probably the correct antiferromagnetically ordered structure has to be taken into
account in the band-structure calculation. In addition, in the existing experiments
only the main branches could be unambiguously assigned.\cite{shishido02}
In Table \ref{table5} we therefore
compare only the experimental and theoretical results for
these main branches. 

\begin{table}[ht]
\caption{Comparison of theoretical and experimental results for the dominating
dHvA frequencies in CeRhIn$_5$. In our calculations (denoted `present') the 
Ce $4f$ electrons are 
%pmo is 
treated either as delocalized ($4f$-val.) or localized
($4f$-c).}
%\begin{tabular}{crrcrrrr} \hline \hline
\begin{ruledtabular}
\begin{tabular}{cccccccc} 
%p \hline \hline
       &       & \multicolumn{3}{c}{theory $F$ (kT)}&\multicolumn{3}{c}{exp.\ $F$ (kT)}\\
       \cline{3-5}\cline{6-8}
symbol  &symbol
&\multicolumn{2}{c}{present }&  & & &  \\ 
\cline{3-4}
%p \cline{3-5}\cline{6-8}
$a$ & $b$ 
&$4f$-val. &$4f$-c	& $b$                          
& $a$ & $b$  & $c$  \\ \hline 
\\
$\beta{_2}  $   	&$F^\prime_8$	&6.53   	&6.07	&6.878
&6.13   	&6.12   	&6.256\\ 
$\alpha{_1}$    	&               	&5.38   	&4.57	&5.562
&4.67   	&	&4.686\\ 
$\alpha{_2}$    	&$F^\prime_7$   	&4.21   	&3.93	&4.331
&3.67   	&3.60   	&3.605\\ 
$\alpha{_3}$    	&$F^\prime_7$   	& 3.90  	&3.48	&3.951
&3.67   	&3.60   	&3.605
\end{tabular}
\end{ruledtabular}
\label{table5}
% \flushleft
{$a$: Ref.\ \onlinecite{shishido02}; $b$: Ref.\ \onlinecite{hall01}; $c$: Ref.\ 
\onlinecite{cornelius00}.}
\end{table}

If we would start our interpretation of the dHvA frequencies in CeRhIn$_5$
with treating 
the Ce $4f$ electrons as valence electrons in the fully relativistic scheme,
then we immediately observe an important difference with regard to both
CeCoIn$_5$ and CeIrIn$_5$: whereas the
experimental data are slightly larger (up to 10 percent) than the theoretical
values for the later two superconducting compounds, they are up to 10 percent
{\it smaller} for CeRhIn$_5$. In contrast, the results obtained with the 
$4f$-c scheme fit much better in the
general chemical trend. Since there is no 4$f$ contribution at the Fermi level
the volume enclosed by the Fermi surface shrinks and so does, consequently,
the corresponding areas of extremal orbits. As a result the theoretical values
are now {\em
smaller} than the measured ones in accordance with the general trend in the
series Ce$M$In$_5$. It should be noted, however, that the absolute deviation
from the experimental values is not significantly improved. 

%pmo - with small changes
What concerns the effective masses, these are enhanced in CeRhIn$_5$ as well,
but only moderately. The experimental value for the mass $m/m_0$ of orbit $\beta_2$
is $6.1 \pm 0.3$ (Ref.\ \onlinecite{hall01}) or $6.5 \pm 0.8$ (Ref.\
\onlinecite{cornelius00}), respectively, whereas theoretically it was found to
be 1.48 (in the fully relativistic scheme) or  0.55 (in the $4f$-c scheme), see
Table \ref{table3}). The corresponding values for orbit $\alpha_2$ are:
experiment: $4.6 \pm 1.0$ (Ref.\ \onlinecite{hall01}) or $4.8 \pm 0.4$ (Ref.\
\onlinecite{cornelius00}), and theory: 0.95 (rel.) or 0.47 ($4f$-c). 
Although the inclusion of $4f$ orbitals into the valence-band complex
enhances the effective masses even in LSDA, that does not necessarily advocate
delocalized $4f$'s. 
%pmo seems not to be the correct enhancement mechanism. 
Instead, it can be expected that the dominating mass enhancement in CeRhIn$_5$ 
occurs because of a strong interaction between conduction electrons
and magnetic fluctuations, which is not taken into
account in our calculation, however. 
Under pressure the antiferromagnetic order disappears in CeRhIn$_5$
yet the spin fluctuations and their interaction with the conduction
electrons intensify, as suggested by the pronounced increase observed for
the effective dHvA masses.\cite{shishido02b,shishido03}

\begin{figure}
\includegraphics{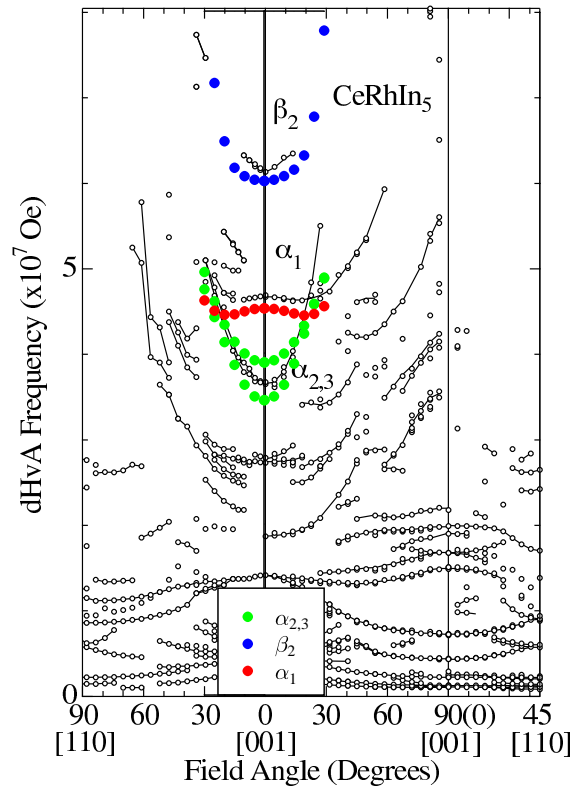}
\caption{(Color online)
Measured angular dependence of dHvA frequencies in CeRhIn$_5$ (open
points, from Ref.\ \onlinecite{shishido02}) in comparison to the theoretical
ones, calculated with  the fully relativistic FPLO scheme and 
treating the Ce $4f$ electron as core electron (filled
symbols).} 
\label{fig3}
\end{figure}

To confirm our interpretation
%pmo 
of localized $4f$ behavior, we calculated in addition the angular dependence  
of the dHvA frequencies of the main branches treating the 4$f$ electrons as 
core states.  The results are shown in Fig. \ref{fig3}. 
As can be seen, the characteristic angular dependence of the various branches is well
reproduced. The calculated splitting of the $\alpha_2$ and $\alpha_3$ branches 
is larger than the experimentally observed splitting,
%pmo which 
the reason why remains an open question at the moment. The related compound LaRhIn$_5$
displayed also a larger $\alpha_2 - \alpha_3$ splitting.\cite{shishido02}
If we
would have displayed the results calculated treating the $4f$ electrons as 
itinerant, then the theoretical branches would lie much higher than the experimental 
ones and the agreement would be much worse. 
Thus, our calculations support the picture of reasonably  localized $4f$ behavior
in CeRhIn$_5$, at variance with the much more delocalized $4f$ behavior
in CeCoIn$_5$ and CeIrIn$_5$.
In Fig.\ \ref{fig3} one can also see 
the multitude of low-frequency orbits, the origin of which could be resolved
only by band-structure calculations taking the antiferromagnetic ordering into account.

%\begin{figure}[cerh-exp4fc.eps]

\section{Conclusions}

Our investigation of the electronic structures and dHvA quantum oscillations 
in the Ce-115 corroborate the following picture of these materials.
In CeCoIn$_5$ and CeIrIn$_5$ the Ce $4f$ electrons appear to be rather
delocalized. Therefore, treating the $4f$
electrons in these two compounds like itinerant states we do obtain a
good description of the measured dHvA frequencies. However, the experimental
effective masses 
of the extremal orbits are considerably enhanced over the band-structure
effective masses. In spite of the substantial many-particle enhancement 
of the masses in the proximity of the heavy fermion state, it is surprisingly
how well the Fermi surface cross sections are described by the band-structure
calculations. 

The behavior of CeRhIn$_5$ differs from that of CeCoIn$_5$ and CeIrIn$_5$.
In order to provide an explanation of the
main dHvA branches of CeRhIn$_5$,  taking into account also
the chemical trend in the series CeCoIn$_5$--CeIrIn$_5$, we find that 
it is better to treat the Ce $4f$ electrons as localized. 
This finding is in agreement with some of the observed differences between
the Rh-compound and the Co- and Ir-compounds, as, e.g., the lower specific heat
and the antiferromagnetic order of the Rh-115 compound. 
The antiferromagnetic structure leads additionally
to a multitude of low-frequency dHvA
branches that are not detected for the other two compounds, an explanation
of which would require a careful study of the Fermi surface of the 
antiferromagnetically ordered compound.

%pmo - new part
The mass enhancement observed within the Ce-115 series is most likely due to
spin fluctuations as well as charge fluctuations, 
but it comes about in a different manner for CeRhIn$_5$. 
The moderate enhancement of the latter
Ce-115 compound occurs due to spin fluctuations on top of the `frozen' magnetic state.
The appreciable mass enhancement in CeCoIn$_5$ and CeIrIn$_5$, on the contrary,
seems to be related to the more itinerant $4f$ electrons, which have a substantial
weight near the Fermi energy and, in addition to being responsible for the 
fluctuating spin polarization, charge fluctuations of the $4f$'s can now contribute 
to the mass enhancement, also.
%pmo Interactions of the atomic spin polarizations could lead to strong 
%pmo correlations. 

%pmo Our results suggest also two different mechanisms for the mass enhancement in
%    the Ce-115 series. The moderate enhancement which was observed in CeRhIn$_5$
%    might occur due to the interaction with magnetic fluctuations such that the
%    $4f$ electrons remain localized. On the contrary, the strong enhancement in
%p   CeCoIn$_5$ and CeIrIn$_5$ (leading to the heavy fermion behavior) can only be
%    explained if the $4f$ electrons become itinerant, but strongly correlated. In
%    the light of this interpretation it is not clear why the application of
%    pressure up to 2.1 GPa did not lead to a delocalization of the $4f$ electrons
%    although superconductivity has already set in and the masses were strongly
%     
%     of pressure up to 2.1 GPa did not increase the corresponding Fermi
%      surface cross sections although superconductivity has already set in 
%
%    enhanced as observed recently in Ref.\ \onlinecite{shishido02b,shishido03}. 
%    It might be that the applied pressure was simply not strong enough, but it is 
%    also clear that a thorough theoretical treatment of mass enhancement in 
%pmo Ce-115 is still lacking. 

Another mystery is why the deviating behavior occurs for the Rh-compound and not,
for example, for the Ir-compound. 
It was noted earlier that small changes in the lattice parameters of the 
Rh-compound do already lead to a physical behavior very similar to that of CeCoIn$_5$
and CeIrIn$_5$,\cite{hegger00} but this does not provide a microscopic explanation.
The small N\'eel temperature of CeRhIn$_5$ ($T_N=$ 3.8 K) suggests that all three compounds
could be close to antiferromagnetic ordering. This gives an indication of what the
pairing mechanism of the unconventional superconductivity
\cite{movshovich01,izawa01,ormeno02}
could be. Most likely significant antiferromagnetic spin fluctuations are
present in all three compounds at low temperatures, the strength of which depends
delicately on the Ce-interplane distance (i.e., the $c/a$ ratio). 
The $c/a$ ratio apparently also has a pronounced influence on the localization 
degree of the Ce $4f$ states.
If the antiferromagnetic interaction is not strong enough to achieve antiferromagnetic
ordering the next favorable order parameter is superconductivity, which might still
benefit from the antiferromagnetic fluctuations present. The fact that the 
superconducting order parameter has unconventional $d_{x^2-y^2}$-symmetry\cite{izawa01}
might also correspond to an unconventional pairing mechanism, however,
it has been shown that this must not always automatically be so.\cite{oppeneer03}

The discussion of the origin of the unconventional superconductivity also
touches upon the importance of relativistic effects, as claimed in Ref.\ 
\onlinecite{maehira03}. In the later work, substantial differences 
between the Fermi surfaces computed with the scalar- and fully relativistic
approximations were reported. In contrast, we find in our investigation that the
results of the scalar- and fully relativistic schemes are very similar for all
the Ce-115 compounds. In particular, no fourth band crossing the Fermi level
is found. Therefore, there is apparently not a compulsory reason for a fully
relativistic description of the superconductivity in the Ce-115 compounds.
\\

\acknowledgments
This work was supported financially by the Egyptian Ministry of Higher Education
and Scientific Research and 
by the German Sonderforschungsbereich 463, Dresden.

\end{document}